# Inter-valley interactions in Si quantum dots

Doyeol Ahn*

*Department of Electrical and Computer Engineering*

*and*

*Institute of Quantum Information Processing and Systems*

*University of Seoul, Seoul 130-743, Republic of Korea*

## Abstract

In this paper, we studied the inter-valley interactions between the orbital functions associated with multi-valley of silicon (Si) quantum dots. Numerical calculations show that the inter-valley coupling between orbital functions increases rapidly with an applied electric field. We also considered the potential applications to the quantum bit operation utilizing controlled inter-valley interactions. Quantum bits are the multi-valley symmetric and anti-symmetric orbitals. Evolution of these orbitals would be controlled by an external electric field which turns on and off the inter-valley interactions. Estimates of the decoherence time are made for the longitudinal acoustic phonon process. Elementary single and two qubit gates are also proposed.

* e-mail: dahn@uos.ac.kr

# I. Introduction

It is well known that the lowest conduction band of an ideal Si crystal has six equivalent minima of ellipsoidal shape along the [100] direction as shown in figure 1. These ellipsoids are often called valleys and the total wave function of the ground state is obtained from a linear combination of the six wave functions each localized around one of the $\Delta_1$ conduction-band minima. The overlap of wave functions associated with different valleys is assumed to be negligible. In the study of early quantum structures such as n-channel inversion layer on the Si (001) surface, it was found that the broken translation symmetry lifts the six-fold degeneracy into the two-fold degenerate valleys located near the X point in the <001> direction in the k-space and the four-fold degenerate valleys in the direction normal to the surface [1].

In addition, there were experimental observations [2-4] of anomalous structures in the gate-voltage dependence of the conductivity of vicinal planes of Si (100) n-channel inversion layers. It has been suggested that these anomalous structures are caused by the lifting of the two-fold valley degeneracy in the <001> direction as a result of the valley-valley interaction [5,6]. The splitting is proportional to the gradient of the confinement potential normal to the surface [7].

It would be interesting to ask whether the inter-valley coupling is controllable. If this is possible, it will permit us extra degrees of freedom in silicon technology. It can also lead to the potential applications to the silicon based quantum information processing. So far, most of the existing proposals for the solid state quantum bits (qubits) are based on the electron spin confined to the quantum-dots [8,9], coherent quantum state in a Cooper-pair box [10], or the nuclear spins of impurity atoms implanted on the surface of Si [11,12]. For the latter it still remains an experimental challenge to fabricate a structure in which each nuclei can be effectively manipulated. Recently, there have been observations of coherent oscillation of a charge qubit in a III-V double quantum dot [13] and stacked coupled quantum dot structures [14]. These results suggest that the controlled evolution of superposed charge states in the semiconductor quantum dots would be possible. In order to implement the solid state quantum computation, however, it is required to minimize the decoherence effects on the coherent quantum states or qubits [15]. This is actually one of the requirements that must be met to make such devices as good candidates for the building block of quantum computer [16, 17]. These conditions include: (1) a scalable physical system with well defined qubits; (2) the ability to initialize the state of the qubits; (3) long relevant decoherence time, much longer than the gate operation time; (4) "universal" set of quantum gates; and (5) a qubit-specific measurement capability.

Potential drawback of the compound semiconductor charge qubit is relatively short decoherence time and difficulties in fabricating double dots. There would be several merits of a silicon implementation of quantum bits. First of all, the crystal growth and processing technology for Si is quite mature. Secondly, some of the scattering processes which contribute to the decoherence such as intra-valley optical phonon processes are forbidden inherently from the group theoretical considerations in the case of silicon. Especially, only the acoustic phonon and the impurity scatterings are allowed within each ellipsoid for the intra-valley processes [18]. In silicon quantum dots, the situation would be more complicated than the inversion layer. The degeneracy of six valleys would be lifted into lower doublet and higher quartet in each quantization axis because of the differences of the effective masses along each axis.

In this paper, we study the inter-valley interactions between the orbital functions associated with multi-valley of silicon (Si) quantum dots. We also consider its potential applications to the quantum bit operation utilizing controlled inter-valley interactions. To the best of our knowledge, the study of inter-valley transitions in the Si quantum dots, especially related to the quantum computation is new.

## II. Theoretical model

Let's consider a quantum dot of cubic geometry with the z-direction assumed to be along the Si (001) surface. Based on Kohn-Luttinger effective mass theory [19], the envelope function for the quantum state in a Si quantum dot is given by

$$F(\vec{r}) = \sum_{\vec{k}} F(\vec{k}) \exp(i\vec{k} \cdot \vec{r}), \tag{1}$$

and

$$F(\vec{k}) = \sum_i \alpha_i F_i(\vec{k}), \tag{2}$$

where $F_i(\vec{k})$ is centered about the $i$ th minimum. The constants $\alpha_i$ can be determined from the group theoretical considerations [20-22]. The equation of motion for $F_i(\vec{k})$ becomes [23,24]

$$\varepsilon_i(\vec{k}) F_i(\vec{k}) + \sum_j \sum_{\vec{k}'} D^{ij}_{\vec{k},\vec{k}'} V(\vec{k} - \vec{k}') F_j(\vec{k}') = \varepsilon F_i(\vec{k}), \tag{3}$$

where $\varepsilon_i(\vec{k})$ is the energy dispersion relation of the $i$-th valley, $V(\vec{k})$ the Fourier component of the total potential, and $D^{ij}_{kk'}$ is the inter-valley coupling term which can be derived from the cell periodic function for the conduction band as

$$\begin{aligned} D^{ij}_{kk'} &= D^{ij}_{\vec{K}_i + \vec{\kappa}, \vec{K}_j + \vec{\kappa}'} \\ &\cong D^{ij}_{\vec{K}_i, \vec{K}_j} + \vec{\kappa} \cdot \frac{\partial}{\partial \vec{K}_i} D^{ij}_{\vec{K}_i, \vec{K}_j} + \vec{\kappa}' \cdot \frac{\partial}{\partial \vec{K}_j} D^{ij}_{\vec{K}_i, \vec{K}_j} \\ &= I_{ij} + \vec{\kappa} \cdot \vec{J}_{ij} + \vec{\kappa}' \cdot \vec{J}_{ij}' \end{aligned} \tag{4}$$

where $\vec{K}_i$ is the wave vector at the minimum at the $i$-th valley. Then within the frame of multi-valley effective mass theory [23,24], the equation of motion for $F_l(\vec{r}) = \sum_{\vec{k}} F_l(\vec{k}) \exp(i\vec{k} \cdot \vec{r})$ can be written down as

$$\left[H_l(-i\vec{\nabla}) + V(\vec{r}) - E\right]F_l(\vec{r}) + \sum_{l' \neq l} H_{ll'}(\vec{r}, -i\vec{\nabla})F_{l'}(\vec{r}) = 0. \tag{5}$$

Here,

$$H_l(-i\vec{\nabla}) = -\frac{\hbar^2}{2m_x}\frac{\partial^2}{\partial x^2} - \frac{\hbar^2}{2m_y}\frac{\partial^2}{\partial y^2} - \frac{\hbar^2}{2m_z}\frac{\partial^2}{\partial z^2} - \frac{ie\hbar B}{2m_x}y\frac{\partial}{\partial x} - \frac{ie\hbar B}{2m_y}x\frac{\partial}{\partial y}$$
$$+ \frac{e^2 B^2}{8}\left(\frac{x^2}{m_y} + \frac{y^2}{m_x}\right) , \tag{6}$$

and

$$H_{ll'}(\vec{r}, -i\vec{\nabla})$$
$$= I_{ll'} \exp[-i(\vec{K}_l - \vec{K}_{l'})\cdot\vec{r}](V(\vec{r}))$$
$$-i(\vec{J}_{ll'} \cdot \vec{\nabla})\exp[-i(\vec{K}_l - \vec{K}_{l'})\cdot\vec{r}](V(\vec{r})) \tag{7}$$
$$+ \exp[-i(\vec{K}_l - \vec{K}_{l'})\cdot\vec{r}](V(\vec{r}))(-i\vec{J'}_{ll'} \cdot \vec{\nabla})$$

and $V(\vec{r}) = V_c(\vec{r}) + e\vec{F}\cdot\vec{r},$ (8)

where $m_x, m_y, m_z$ are effective masses along x, y, z directions in each valley, $E$ is quantized energy, $\vec{K}_l$ is the wave vector at the minimum at the $l$-th valley, $I_{ll'}, \vec{J}_{ll'}, \vec{J'}_{ll'}$ are inter-valley coupling terms, $V_c(\vec{r})$ is the quantum dot confinement potential, and $\vec{F}$ is an applied electric field.

In order to calculate the inter-valley coupling terms, we assume that $D^{ll'}_{\vec{K}_l, \vec{K}_{l'}}$ can be expressed by the following simple form [24]

$$D^{ll'}_{\vec{K}_l, \vec{K}_{l'}} = I_{ll'} = \alpha \vec{e}_l \cdot \vec{e}_{l'} + \beta, \tag{9}$$

where $\vec{e}_l$ is the unit vector in the direction of $l$-th axis and $\alpha, \beta$ are constants to be determined from the band-structure parameters. For example, Cardona and Pollak [25] gave

$$D^{13}_{(K,0,0),(0,K,0)} = 0.3915, D^{12}_{(K,0,0),(-K,0,0)} = -0.2171 \tag{10}$$

with $K = 0.85 \times 2\pi/a$ and the lattice constant $a = 0.543 nm$ for Si. On the other hand, Shindo and Nara [23] gave slightly different numbers. From equations (9) and (10), we have

$$I_{(K_0,0,0)(0,K_0,0)} = \beta = 0.3915,$$
$$I_{(K_0,0,0)(-K_0,0,0)} = -\alpha + 0.3915 = -0.2171, \tag{11}$$

which give $\alpha = 0.6086$. Then from equations (4), (9)-(11), we obtain

$$I_{ll'} = \frac{1}{2}(1 + \vec{e}_l \cdot \vec{e}_{l'}) - \frac{1}{2}(1 - \vec{e}_l \cdot \vec{e}_{l'})\cos(2\lambda_K), \tag{12}$$

and

$$\vec{J}_{ll'} = \frac{\partial}{\partial \vec{K}_l} I_{ll'}$$
$$= \vec{e}_l \frac{\partial}{\partial K} I_{ll'}$$
$$= \vec{e}_l (1 - \vec{e}_l \cdot \vec{e}_{l'}) \frac{\partial \lambda_K}{\partial K} \sin(2\lambda_K) \tag{13}$$

with $\quad \tan(2\lambda_K) = \dfrac{2TK}{\varepsilon_G}$, $\tag{14}$

where T=1.08 a.u. and $\varepsilon_G = 0.268 Ry$. Here, we have included only the bases of $\Gamma_1^u$ and $\Gamma_{15}$ in the representation.

The most important feature of our model is that the inter-valley coupling can be turned on and off by the applied electric field. For example, the inter-valley coupling between the valley 5 and the valley 6 (along z-axis) is approximated by [24]

$$H_{56} = -I_{56} \exp[-i(\vec{K}_5 - \vec{K}_6) \cdot \vec{r}](V_c(\vec{r}) + eFz)$$

$$-i|J_{56}| \frac{\partial}{\partial z} \exp[-i(\vec{K}_5 - \vec{K}_6) \cdot \vec{r}](V_c(\vec{r}) + eFz) \tag{15}$$

$$+ \exp[-i(\vec{K}_5 - \vec{K}_6) \cdot \vec{r}](V_c(\vec{r}) + eFz)\left(-i|J_{56}| \frac{\partial}{\partial z}\right),$$

with

$$I_{56} = -\cos(2\lambda_K),$$
$$J_{56} = J'_{56} = 2\frac{\partial \lambda_K}{\partial K} \sin(2\lambda_K). \tag{16}$$

Here, we substituted equations (12) and (13) into equation (7) for $l = 5$ and $l' = 6$ and assumed that the electric field $F$ is in the z-direction.

**III. Numerical Results and Discussions**

We have solved equations (5) to (16) for the Si quantum dot structure mentioned above numerically. We also considered potential quantum bit operation utilizing the inter-valley interactions. In this work, we considered a quantum dot with the dimension of 8nm, 12nm, and 6nm in x-, y- and z-directions, respectively. In this structure, the ground state is associated with doubly degenerate valleys 5 and 6.

When the weak static magnetic field is applied along the growth direction, the ground state wave function is composed of the linear combination of p-like $T_1$ states from the irreducible representations of $T_d$ symmetry of the Si crystal [26]. In other word, the ground state wave function is given by $|\Psi\rangle = \frac{1}{\sqrt{2}}(|F_5\rangle \pm |F_6\rangle)$, where $F_5$ and $F_6$ are orbital functions for the valley 5 and 6, respectively. These orbitals satisfy the following effective Hamiltonian in the interaction picture:

$$H = \begin{bmatrix} \varepsilon(F) & \Delta(F) \\ \Delta(F) & \varepsilon(F) \end{bmatrix}. \qquad (17)$$

Here $\varepsilon$ is the energy difference between symmetric and anti-symmetric states, $\Delta$ is the inter-valley coupling, and $F$ is an external electric field along the z-direction. When $F = 0$, both $\varepsilon$ and $\Delta$ are zero and the total state remains unchanged because there is no inter-valley coupling. In this model, we have neglected the coupling of orbitals between different axes. For example, the coupling between valleys 1 and 5 (x-axis and z-axis) is found to be million times smaller than the coupling between the valleys 5 and 6 (both in the z-axis). If we apply an external electric field to the quantum dot, the inter-valley interaction is turned on and doubly degenerate ground state is splitted. The crystal momentum necessary for the coupling of electron states between the valley 5 and the valley 6 is provided by an applied electric field along the z-direction [18].

In Fig.2, we plot the energy difference $\varepsilon$ between the symmetric and the anti-symmetric states and the inter-valley coupling energy $\Delta$ which is defined as $\Delta(F) = <F_5 | H'_{56} | F_6>$. In this figure, one can see that the inter-valley coupling is increasing rapidly with the electric field. For example, the calculated values of $\varepsilon$ and $\Delta$ are $63.5\,\mu eV$ and $31.6\,\mu eV$, respectively, when $F = 400\,kV/cm$. When F is increased to 500 kV/cm, we have $\Delta = 43\mu eV$. These field values are below the breakdown field strength as can be seen by the recent experimental results for the inversion layer mobility which has been measured up to 1MV/cm [27].

If we turn on the electric field and wait long enough, the system would be in the symmetric state which is denoted by |0>. The coherent evolution from the symmetric state |0> to the anti-symmetric state |1> would be observed by applying the sharp voltage pulse to the pulse gate similar to the case of the Cooper-pair box [10] or the double quantum dot structure [13,14]. The coherent oscillation of the system is

characterized by the angular frequency given by $\Omega = \sqrt{\varepsilon^2 + \Delta^2}/\hbar$, which corresponds to the microwave frequency of 17.2GHz. When the system is evolved to the state |1> and if we turn off the electric field $F$ adiabatically, the inter-valley coupling would be turned off. The resulting state would be the anti-symmetric orbitals which would maintain its phase coherence until the decoherence destroys the coherence.

Figure 3 shows the first 6 energy levels associated with valley 5 (or 6) in solid lines, valley 1 (or 2) in dashed lines, and valley 3 (or 4) in dotted line, as functions of increasing electric field. Weak magnetic field of 1.5 Tesla is applied along the z-axis. The dimension of the quantum dot used in this particular calculation is such that the ground state is associated with valley 5 or 6 in the absence of an external field. It is interesting to note that the slopes for energy levels associated with the valleys 1 and 3 are similar but they are different from those of the valley 5 because of the effective mass differences along the field direction. The energy states are labeled for the single valley case, that is, when the intervalley coupling is ignored. Part of the ground state energy level is magnified and shown in the small box inside the figure 3. One can notice that the ground state energy is further splitted into symmetric and anti-symmetric states. It is interesting to see that $E_3$ and $E_5$ associated with valleys 5 and 6 show anti-crossing at point D with increasing electric field. The inset shows the magnification of point D.

Details of anti-crossing behavior is shown in Fig. 4 for the symmetric states (solid lines) and anti-symmetric states (dashed lines) associated with $E_3$ and $E_5$, respectively. We found that anti-crossing occurs at the field strength of 131.6 kV/cm and the energy gap is $117\,\mu eV$. At low electric field, $E_3$ is pushed up while $E_5$ is showing the negative shift with increasing electric field until anti-crossing point D and their behaviors are changed the other way around after passing D. Similar behavior was observed in the case of quantum well with an applied electric field [28].

The symmetric and anti-symmetric splitting and other abundant features of the energy level spectrum of Fig. 3 open up strong possibilities of realizing orbital qubits and quantum gates. The simplest example would be the controlled electric field induced transition between symmetric and anti-symmetric states associated with valleys 5 and 6. The insets of Fig. 3 shows a magnified energy diagrams. We first consider the symmetric and anti-symmetric states associated with $E_0$ (point C). Initially, we set the electric field at a low value (point A) so that the transition between two states is difficult to occur (Fig. 2) due to a relatively small $\Delta$. The electron in the quantum dot is in the ground state. When the gate bias is switched to a higher electric field (point B), the time evolution between two states would begin. The time interval of the pulse determines the relative population of two states and they would remain at the final values when the pulse is switched back to A. The rise time of the pulse should be shorter than $\hbar/\Delta$ at A and longer than $\hbar/\Delta$ at B. On the other hand, one can also utilize the anti-crossing for qubit operation shown in Fig. 4 for a qubit operation, following similar approach for the superconducting qubit [10]. Qubit is prepared at E (Fig. 4) by charging an electron into the anti-symmetric state associated with $E_3$. We increase the electric field adiabatically to the point F and then apply the microwave to start the qubit operation. The read-out can be done by decreasing the electric field adiabatically to point E again. The read-out of the relative population can be achieved by measuring the transport through quantum dot. Since it is important to control both the potential and the electric field across the quantum dot, the biases of all terminals (source, drain, front gate, back gate) should be adequately adjusted.

When the ground state is associated with valley 5 and valley 6 only, the wave function can be written as

$$\Psi_{S,A} = \frac{1}{\sqrt{2}} \begin{pmatrix} 1 \\ \pm 1 \end{pmatrix} \varphi_{S,A}(\vec{r}) = \chi_{S,A} \varphi_{S,A}(\vec{r}), \tag{18}$$

where $\varphi_{S,A}(\vec{r})$ are the orbital wave functions and $\chi_{S,A}$ are the pseudo-spins for the symmetric and anti-symmetric state, respectively.

Fig. 5 shows the schematic of the single qubit operation and the read-out circuit. One can use a silicon-on-insulator (SOI) quantum dot structure for qubit operation. The quantum dots are surrounded by $SiO_2$ and two independently tunable gates are formed on top of $SiO_2$. The biases on the center gates ($V_{G1}$ and $V_{G2}$) and the back gate (or the ground plane) are tuned such that the required electric field in the Si quantum dot is generated in the z direction. For F = 300 ~ 500 KV/cm, the quantum dot should be in the sub-threshold region. The biases on the left and the right gates ($V_S$ and $V_D$) induce the tunneling of an electron from the dot 1 to dot 2 during the read-out. The value $V_D - V_S$ is kept smaller than $k_B T$ so that the QD is in the linear transport regime. In this bias scheme, F is large only in the z direction. The quantum state of the single electron injected into the quantum dot 1 is the qubit and the quantum dot 2 which is coupled to dot 1, acts as a read-out device. The tunneling probability amplitude between dot 1 and dot 2 is proportional to [29]

$$T_{12} = \iint d\vec{r}_1 d\vec{r}_2 \varphi_a^1{}^*(\vec{r}_1) H_T(\vec{r}_1,\vec{r}_2) \varphi_b^2(\vec{r}_2)(\chi_a^1)^\dagger \chi_b^2, \tag{19}$$

where $a,b = S$ or $A$ and $H_T$ is the tunneling Hamiltonian. It is interesting to note that the quantum mechanical tunneling of an electron between quantum dot 1 and dot 2 is parity dependent. In other words, the tunneling probability is non negligible when the initial and the final states are in the same parity states, either symmetric or anti-symmetric. The $\sigma_x$ operation on the qubit is achieved by the gate voltage $V_{G1}$ and the microwave pulse. Measurement of the qubit (quantum dot 1) can be done by adjusting the gate voltage $V_{G2}$ such that the ground state of quantum dot 2 is in resonance with the symmetric state of a dot 1 while the anti-symmetric states are off resonant, and by setting $V_D - V_S$ to induce the tunneling. We can also design the quantum dot 2 to meet this condition. Since the energy gap $2\Delta$ between $|1\rangle$ and $|0\rangle$ is an order of 50 $\mu V$, the ambient temperature around 30 mK would be required to suppress the decoherence. From equation (19), one can see that the tunneling probability of the symmetric state ($|0\rangle$) would be larger than the anti-symmetric state ($|1\rangle$) due to the parity selection rule. The presence or absence of an excess electron in a quantum dot 2 will be denoted as the logical state $|0_L\rangle$ or $|1_L\rangle$, respectively. The excess charge of a dot 2 due to the tunneling process can be measured using sensitive single electron capacitance technique [30-32].

We now consider the implementation of a non-trivial two-qubit gate. In Fig. 6, we show the elementary two-qubit quantum gate, which is comprised of four quantum dots, two for the two qubits and the rest for the read-out. Quantum dots 1 and 2 are coupled by inter-dot Coulomb interaction which is also parity dependent. The inter-dot Coulomb interaction energy is calculated by following Beattie and Landsberg [33]:

$$V_{if} = \iint d\vec{r}_1 d\vec{r}_2 [\varphi_1*(\vec{r}_1)\varphi_2*(\vec{r}_2)\Delta_{21} - \varphi_2*(\vec{r}_1)\varphi_1*(\vec{r}_2)\Delta_{12}] \\ \times V_{sc}(\vec{r}_1 - \vec{r}_2)[\varphi_1(\vec{r}_1)\varphi_2(\vec{r}_2)] \qquad (20)$$

where $V_{sc}$ is the screened Coulomb potential; $\Delta_{21} = \Delta_{12} = 1$ when electrons in dot 1 and dot 2 have the same parities; $\Delta_{21} = 1$ and $\Delta_{12} = 0$ when electrons 1 dot 1 and 2 have the opposite parities, which are preserved; and $\Delta_{21} = 0$ and $\Delta_{12} = 1$ when electrons 1 dot 1 and 2 have the opposite parities, which are both changed. The Hamiltonian for this two electron system is given by ( in $|11\rangle, |10\rangle, |01\rangle, |00\rangle$ bases):

$$\hat{H} = \begin{bmatrix} E_{11} & 0 & 0 & 0 \\ 0 & E_{10} & E_c & 0 \\ 0 & E_C & E_{01} & 0 \\ 0 & 0 & 0 & E_{00} \end{bmatrix}. \qquad (21)$$

Let's consider the special case of $E_{11} = 3\Delta$, $E_{10} = E_{01} = \Delta$, $E_{00} = -\Delta$ and $E_C = \delta$, and let the system evolves unitarily for the time t. The unitary evolution operator is given by

$$\hat{U} = \exp(i\hat{H}t)$$

$$= \exp(3i\Delta t)|11\rangle\langle 11| + \left(\cos\Omega_1 t + i\frac{\Delta}{\Omega_2}\sin\Omega_2 t\right)(|10\rangle\langle 10| + |01\rangle\langle 01|) \quad (22)$$

$$+ \exp(-i\Delta t)|00\rangle\langle 00| + \left(\cos\Omega_3 t - 1 + i\frac{\delta}{\Omega_2}\sin\Omega_2 t\right)(|10\rangle\langle 01| + |01\rangle\langle 10|),$$

where $\Omega_1 = \sqrt{\Delta^2 + \delta^2}$, $\Omega_2 = \sqrt{\Delta^2 + 3\delta^2}$, and $\Omega_3 = \sqrt{2}\delta\Delta$. The last term in the equation (22) describes the swap operation $|10\rangle \to |01\rangle$ and vice versa. If the initial state is $|10\rangle$, the resulting state after the unitary evolution for the time t will become

$$|10\rangle \to (\cos\Omega_1 t + i\frac{\Delta}{\Omega_2}\sin\Omega_2 t)|10\rangle + (-1 + \cos\Omega_3 t + i\frac{\delta}{\Omega_2}\sin\Omega_2 t)|01\rangle. \quad (23)$$

If we set $t = \pi/(2\Omega_3)$ and $\Delta = (4 + \sqrt{13})\delta$, we get

$$|10\rangle \to -|01\rangle + \cos\left(\frac{\pi\Omega_1}{2\Omega_2}\right)|10\rangle \approx -|01\rangle, \quad (24)$$

which is a swap operation followed by the phase change. In order to synthesize the controlled NOT (CNOT) operation, we need to supplement the one qubit operation to the above operation. In passing, we would like to comment that our proposal is based on adiabatic switching of an electric field and is expected to be quite slow.

Once the valley interaction is turned off, the quantum state is supposed to evolve unitarily until the decoherence processes destroy the coherence [34,35]. Since both $F_5$ and $F_6$ are in ground states, respectively, the only coherency destroyed by the decoherences is their relative phase. Here, we estimate the phase decoherence by the longitudinal acoustic (LA) phonons. The upper bound of the scattering rate due to the LA phonon is given by

$$W^\pm = \frac{2\pi}{\hbar}\sum_f \sum_{\vec{q}} E_{ac}^2 \frac{\hbar q^2}{2V\rho\omega_q}(N_{\vec{q}} + \frac{1}{2} \pm \frac{1}{2})|<f|e^{\mp i\vec{q}\cdot\vec{r}}|i>|^2 \delta(E_f - E_i \mp \hbar\omega_q)$$

$$\leq \frac{2\pi}{\hbar} \sum_f \sum_{\vec{q}} E_{ac}^2 \frac{\hbar q^2}{2V\rho\omega_q} (N_{\vec{q}} + \frac{1}{2} \pm \frac{1}{2}) \delta(E_f - E_i \mp \hbar\omega_q)$$
$$\approx 4\pi^2 \frac{(E_f - E_i)^3 E_{ac}^2}{\rho \hbar^4 c_l^5} \exp(-(E_f - E_i)/k_B T)$$
, (17)

where $\rho = 2.33 (g/cm^3)$, $c_l = 9.01 \times 10^5 (cm/\sec)$, and $E_{ac} = 4.7 (eV)$ for Si. For more detailed calculations of phonon scattering, we refer the work of Fischetti and Laux [36]. In Fig. 7 (a), we show the lower bounds of the intra-valley relaxation times (or the upper bounds of the scattering rates) for different energy fluctuations as functions of the lattice temperature. In quantum dots, the phonon scattering rates are considerably lower than those of the bulk or the quantum wells because only the transitions between discrete states are allowed. Fig. 7 (b) shows the estimates of decoherence time (or intra-valley relaxation time) due to the LA phonons for different lattice temperatures as functions of the fluctuation energy. Both figures indicate the decoherence time of an order of 100 nanosecond to microsecond for Si quantum dot structures, which is considerably longer than the III-V quantum dots.

The de-phasing time (or decoherence time) of the spin qubit for bulk GaAs or GaAs quantum dot is an order of microsecond [37,38], whereas the decohrence time for the charge qubit is less than nanosecond [9,39]. The estimated decoherence time in Fig. 7 is in the same order of magnitude as that of the spin qubit and much longer than that of the charge qubit. We would also like to emphasize that our case is for the single quantum dot and with the decoherence time comparable to the spin case.

Once the external field is turned on adiabatically, the quantum state will evolve between the symmetric and anti-symmetric states and the operation time would be proportional to $\hbar/\Delta$ which is an order of 0.1 nsec. From this, we expect that about one thousand state evolutions (or operations) would be possible before the decoherence processes destroy the coherence of the quantum state.

**IV. Summary**

In summary, we studied the inter-valley quantum state transitions in a Si quantum dot theoretically. We also investigated the possibility of utilizing these inter-valley transitions for a quantum bit operation. Quantum bits are the multi-valley symmetric and anti-symmetric orbitals. Evolution of these orbitals would be controlled by an external electric field which turns on and off the inter-valley interactions. Estimates of the decoherence time are made for the longitudinal acoustic phonon process. Elementary single and two qubit gates are also proposed.

**Acknowledgements**

This work is supported by the University of Seoul through the 2004 Research Grant. D. A. also thanks D. K. Ferry, S. W. Hwang, S.-H. Park, J. H. Oh and J. H. Lee for valuable discussions.

**Figure Captions**.

Fig. 1  The lowest conduction band of an ideal Si crystal with six equivalent minima of ellipsoidal shape along the [100] direction. For example, $K_5 = (0, 0, 0.85 \times \frac{2\pi}{a})$.

Fig. 2  We plot the energy difference $\varepsilon$ between the symmetric and the anti-symmetric states as well as the inter-valley coupling energy $\Delta$ of a Si quantum dot as functions of the electric field.

Fig. 3  We plot the first 6 energy levels associated with valley 5 (or 6) in solid lines, valley 1 (or 2) in dashed lines, and valley 3 (or 4) in dotted line as functions of increasing electric field.  Weak magnetic field of 1.5 Tesla is applied along the z-axis. The insets of Fig. 3 show a magnified energy diagrams.

Fig. 4  Details of anti-crossing behavior is shown for the symmetric states (solid lines) and anti-symmetric states (dashed lines) associated with $E_3$ and $E_5$, respectively.

Fig. 5  Schematic of qubit operation and read-out.

Fig. 6  Layout for the two quantum bit gate.

Fig. 7  (a) The lower bounds of the intra-valley relaxation times (or the upper bounds of the scattering rates)for Si quantum dot for different energy fluctuations as functions of the lattice temperature due to the LA phonons are plotted.

(b) We show the estimates of decoherence time (or intra-valley relaxation time) for orbital qubit of a Si quantum dot due to the LA phonons for different lattice temperatures as functions of the fluctuation energy.

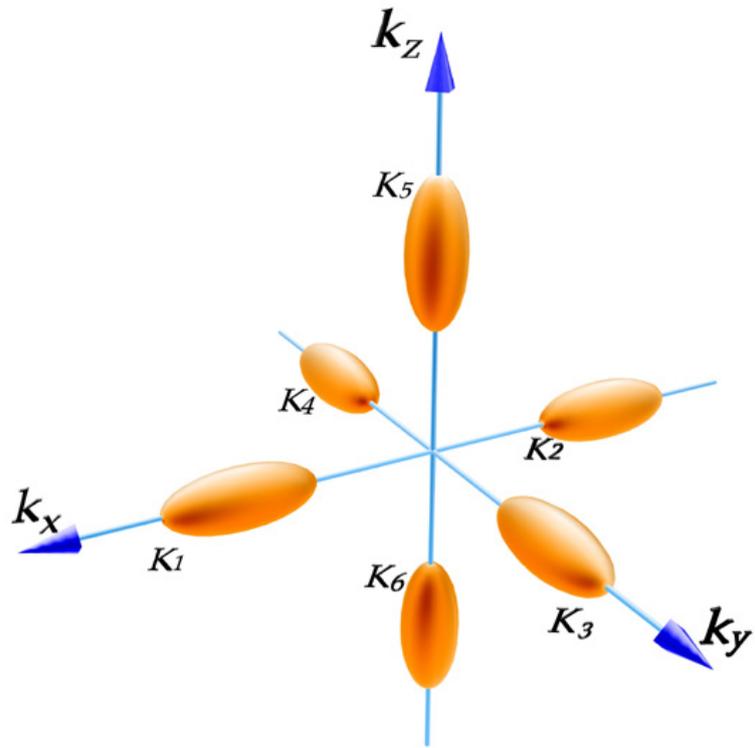

Figure 1

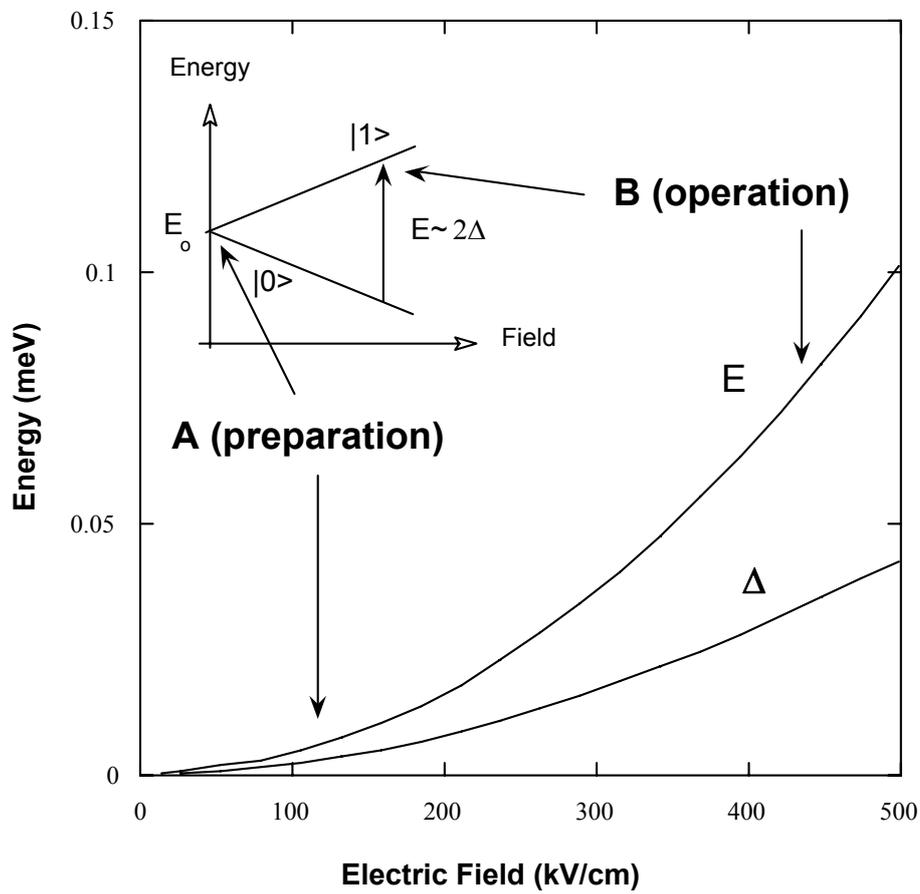

Figure 2

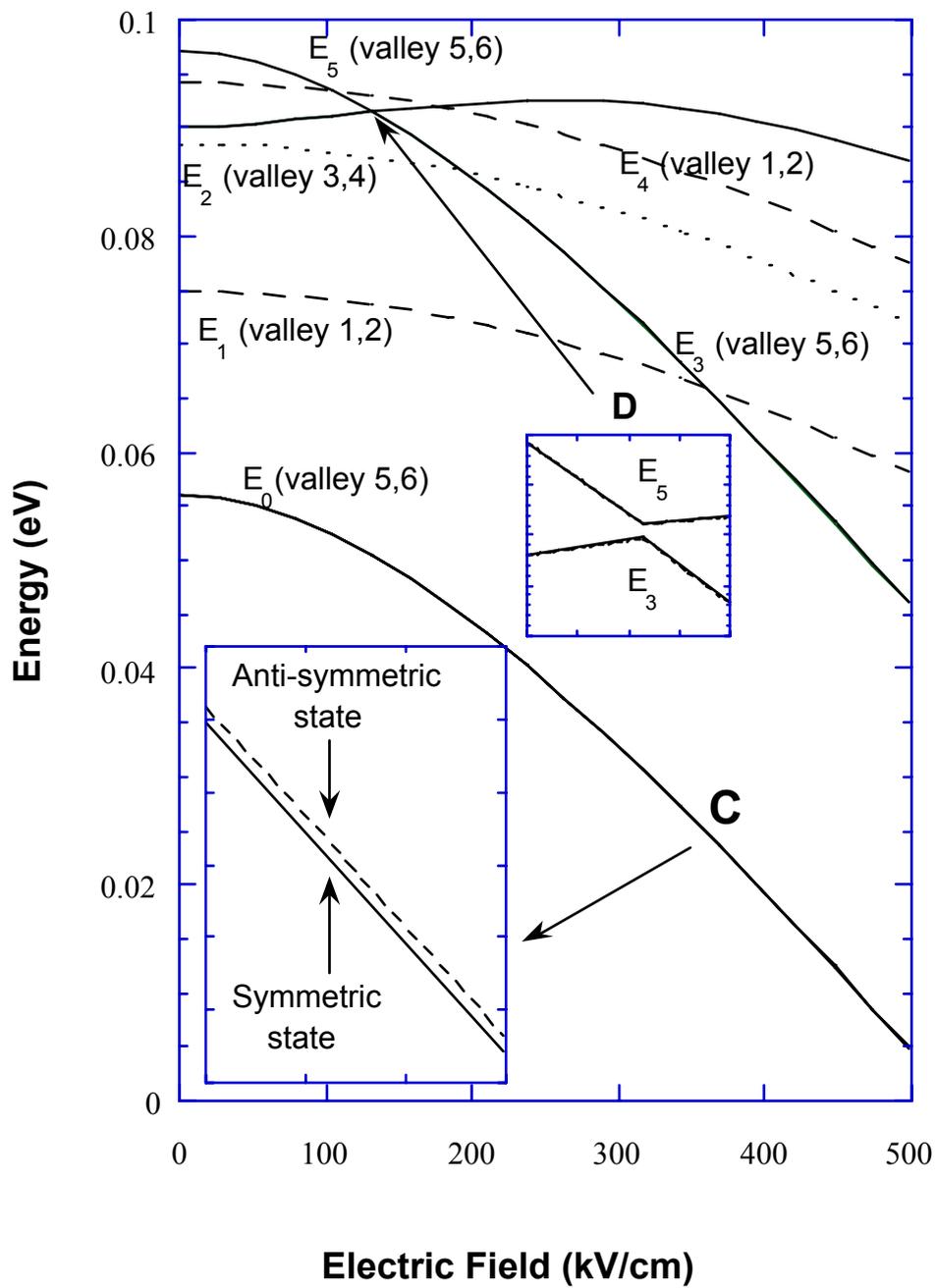

Figure 3

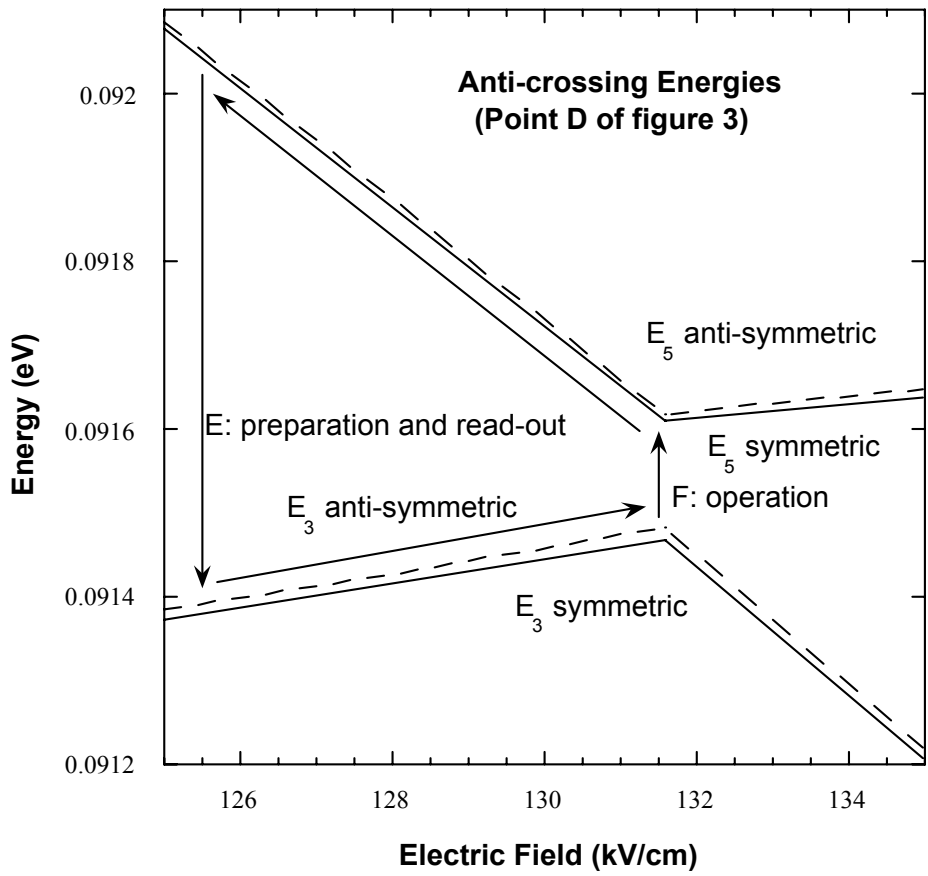

Figure 4

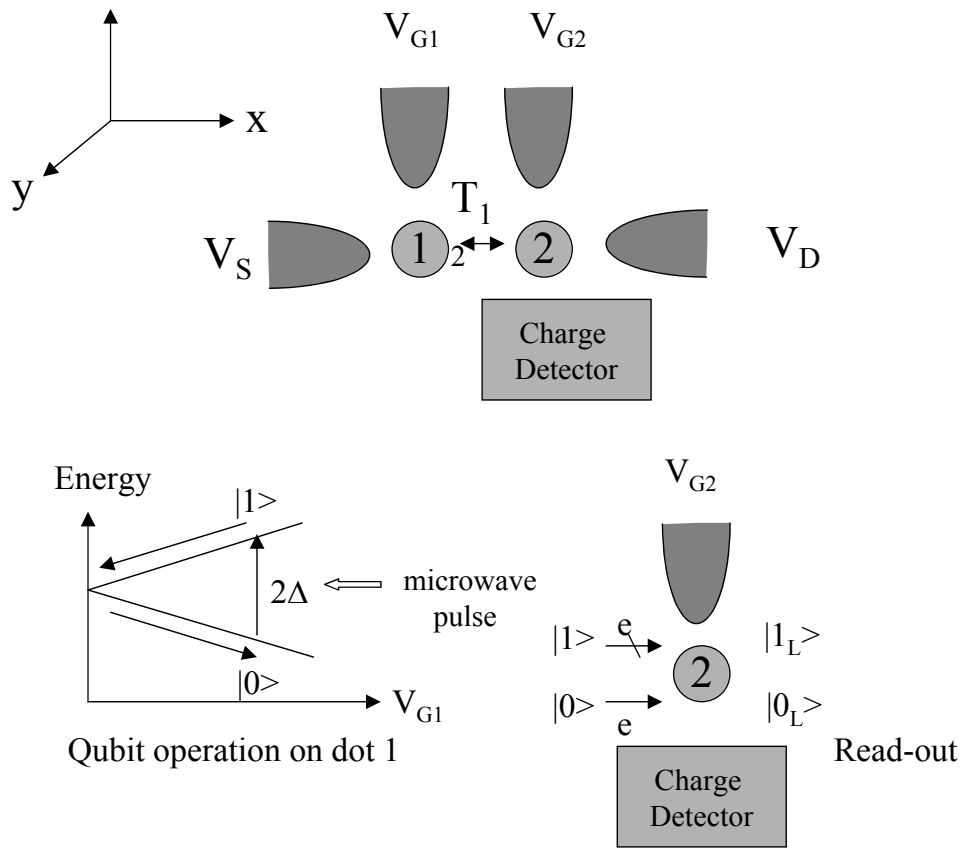

Figure 5

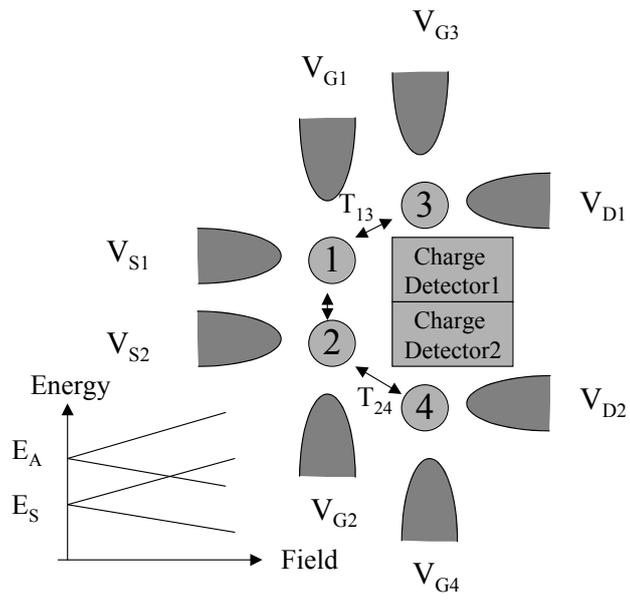

Figure 6

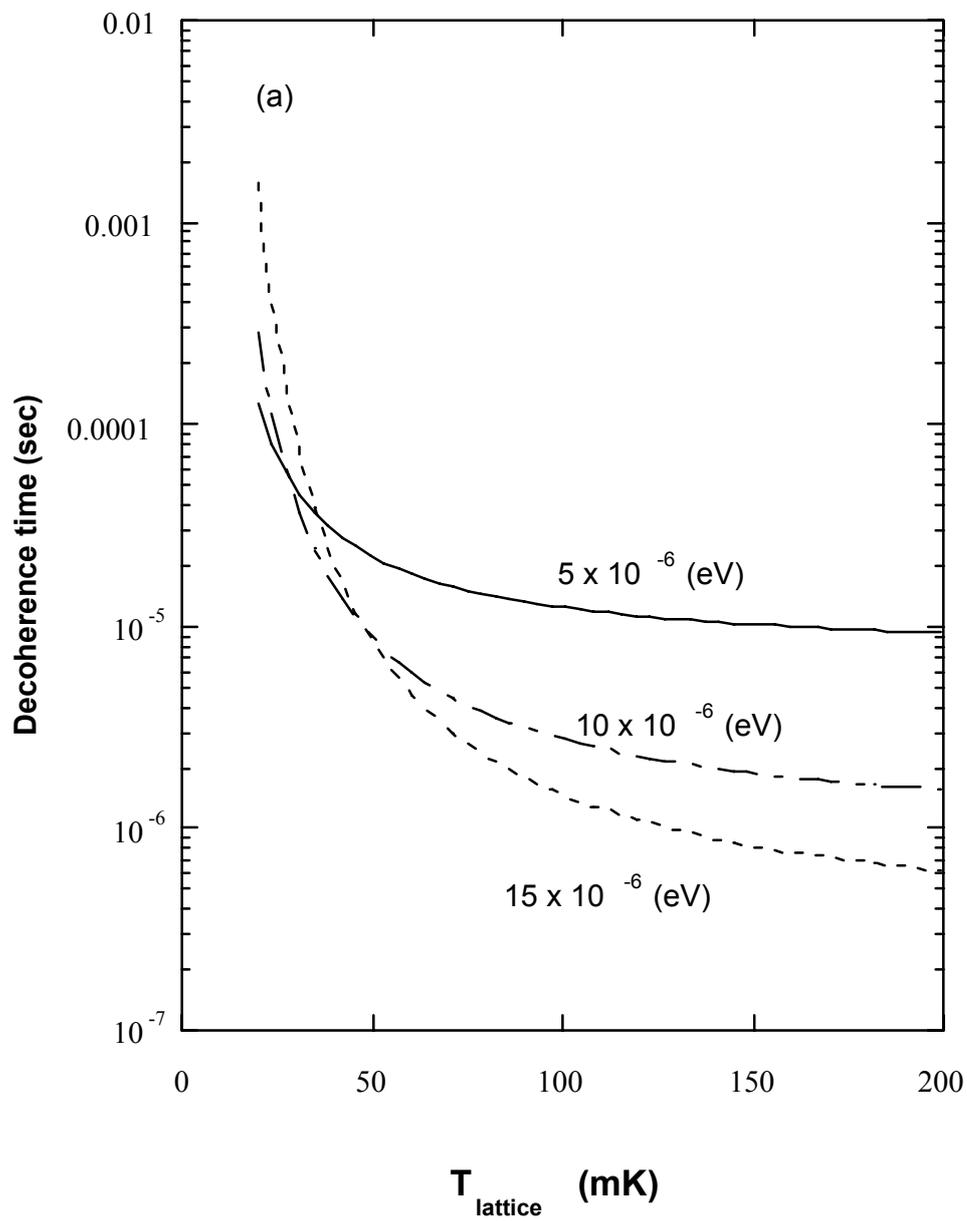

Figure 7 (a)

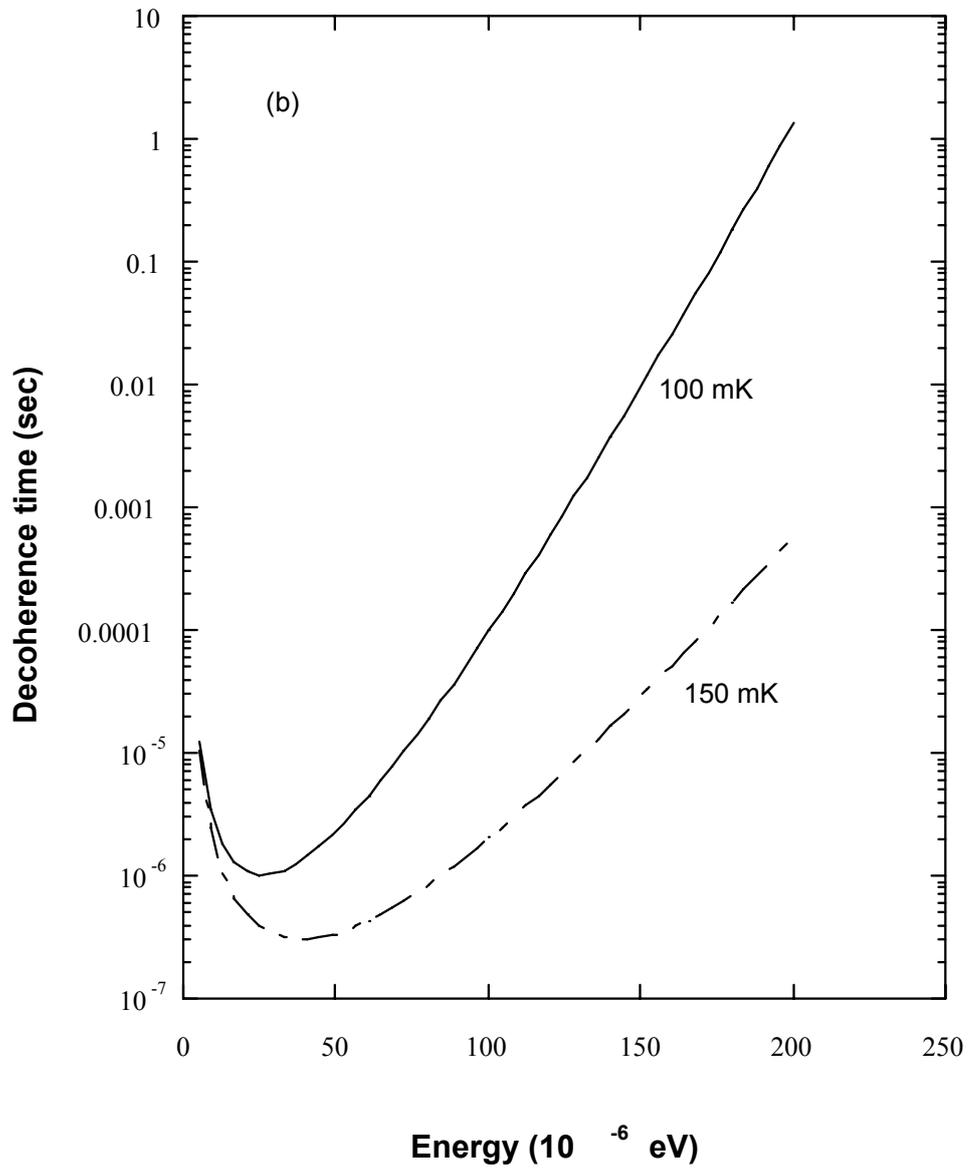

Figure 7 (b)